 \documentclass[final,3p,times]{elsarticle}

\usepackage{amssymb}

\journal{Journal Name}

\newtheorem{thm}{Theorem}
\newtheorem{lem}{Lemma}
\newdefinition{rmk}{Remark}

\newproof{pf}{Proof}
\newproof{pot}{Proof of Theorem \ref{thm2}}

\begin{document}

\begin{frontmatter}

\title{Discriminating between two models based on Bregman divergence in small samples}

\author[mymainaddress,mysecondaryaddress]{Papa NGOM\corref{mycorrespondingauthor}}
\cortext[mycorrespondingauthor]{Corresponding author}
\ead{papa.ngom@ucad.edu.sn}
\author[mysecondaryaddress]{Jean de Dieu NKURUNZIZA}
\author[mysecondaryaddress]{Carlos Simplice OGOUYANDJOU}
\address[mymainaddress]{LMA,Universit\'{e} Cheikh Anta Diop, Dakar, Senegal}
\address[mysecondaryaddress]{Institut de Math\'ematiques et de Sciences Physiques, Porto Novo, Benin}

\begin{abstract}
Recently in \cite{19,20}, Ali-Akbar Bromideh introduced the Kullback-Leibler Divergence (KLD) test statistic in discriminating between two models. It was found that the Ratio Minimized Kulback-Leibler Divergence (RMKLD) works better than the Ratio of Maximized Likelihood (RML) for small sample size. The aim of this paper is to generalize the works of Ali-Akbar Bromideh by proposing a hypothesis testing based on Bregman divergence in order to improve the process of choice of the model. Our aproach differs from him. After observing $ n $ data points of unknown density $ f $;  we firstly measure the closness between the bias reduced  kernel density estimator and the first estimated candidate model. Secondly between the bias reduced  kernel density estimator and the second estimated candidate model. In these two cases Bregman Divergence (BD) and the bias reduced kernel estimator \cite{29} focuses on improving the convergence rates of kernel density estimators are used. Our testing procedure for model selection is thus based on the comparison of the value of model selection test statistic to critical values from a standard normal table. We establish  the asymptotic properties of Bregman divergence estimator and approximations of the power functions are deduced. The multi-step MLE process will be used to estimate the parameters of the models. We explain the applicability of the BD by a real data set and by the data generating process (DGP). The Monte Carlo simulation and then the numerical analysis will be used to interpret the result. 
\end{abstract}

\end{frontmatter}


\section{Introduction}
\label{Intro}
Bregman (1967) introduced for convex functions\cite{21,22,23,24}, the nonnegative measure of dissimilarity. His motivation was the problem of convex programming, but in the subsequent literature it became widely applied in many other problems under the name Bregman distance in spite of that it is not in general the usual metric distance (it is a pseudodistance which is reflexive but neither symmetric nor satisfying the triangle inequality).  In the last decade, Bregman divergences have become an important tool in many research areas. For instance, several specific Bregman divergences, such as Itakura-Saito Distance
 \cite{25,26,27}, Kullback-Leibler Divergence (KLD)
 \cite{28},  and Mahalanobis Distance (P.C. Mahalanobis; 1936) have been used in machine learning as the distortion functions (or loss functions) for clustering tasks. These divergences have been used in generalizations of principal component analysis to data with distributions belonging to the exponential family. Although, the goodness-of-fit and significance testing is initially used in selection of two probability densities; many models selection criteria have been proposed so far. Classical model selection criteria using least square error and log-likelihood include the $C_{p}$-criterion, cross-validation (CV), the Akaike Information Criterion (AIC) based on the well-known Kullback-Leibler divergence, Bayesian Information Criterion (BIC), a general class of criteria that also estimates the Kullback-Leibler Divergence (KLD). These criteria have been proposed by Mallows \cite{9}
 Stone \cite{10}, Akaike \cite{11}, Schwarz \cite{12} and Konishi and Kitagawa \cite{13},
  Aida Toma \cite{37} respectively. Ngom and Ntep \cite{16,17} provided (in their  outcomes of the tests) information on the strength of the statistical evidence for the choice of a model based on its goodness-of-fit. Vuong's tests \cite{18} for model selection rely on the Kullback-Leibler  Information Criterion (KLIC). Mohd Saat et al (2008) compared RML with Vuong's closeness test to discriminant between Gamma and Weibull, in which they found both methods relatively similar. Similar work to compare RML with Kolmogorov-Smirnov and Chi-Squared (with asymptotic properties) has been studied and some inconsistency among them are reported (Basu et al, 2009). Despite of significant amount of work on discrimination by different methods, there is no much work to use the KLD and compare it with alternatives test statistic \cite{19,20}. Akbar introduced the Kullback-Leibler Divergence (KLD) test statistic in discriminating between two models. It was observed that the proposed test statistic named Ratio of minimized Kullback-Leibler divergence (RMKLD)  is consistent with alternative testing statistic, say RML. The purpose of this paper is to generalize the works of Ali-Akbar Bromideh \cite{20}
 by proposing a hypothesis testing based on Bregman divergence in order to improve the process of choice of the model. Our model selection approach differs from him. The testing procedure for model selection will be based on the comparison of the value of Bregman type statistic to critical values from a standard normal table. Following Vuong \cite{14} the procedures considered here are testing the null hypothesis that the competing models are equally close to the data generating process (DGP) versus the alternative hypothesis that one model is closer to the DGP where closeness of a model is measured according to the discrepancy implicit in the Bregman divergence type statistic used.\\ 
 
The rest of the paper is organized as follows. We give some definitions and notations in Section 2. For proving the results rigorously we briefly describe a  reduced bias kernel estimator and the Bregman divergence in Section 3 and 4.  In Section 5, we establish the consistancy of the Bregman divergence estimator. In Section 6 we obtain asymptotic distributions of the Bregman divergence estimator. Model selection and Bregman divergence  based test Statistic are presented in Section 7. Examples and Data Analysis: Implementation of the Bregman divergence test statistic and the probability of correct selection (PCS) is obtained in Section 8 and finally the conclusion appears in Section 9.

\section{Definitions and Notations}\label{sec2}

Let  $ \left( \mathcal{X}, \beta_{\mathcal{X}},F \right)  $ be the statistical space associated with the support $\mathcal{X}=\lbrace 1,2,...,M_{0}\rbrace, ~\forall M_{0}\geq 1 $;~ $ \beta_{\mathcal{X}} $ is the $  \sigma$-algebra defined on $\mathcal{X}$ and $(\mathcal{X},\beta_{\mathcal{X}}) $, the measurable space.

Let  
\begin{eqnarray*}
\Lambda_{M_{0}}=\left\lbrace  F=(f_{1},..., f_{M_{0}})^{T}; \forall x\in \mathbb{R} ~~f_{i}(x)\geq 0, ~i=1,...,M_{0}~~\textsl{and} ~ \sum_{i=1}^{M_{0}}f_{i}(x)=1 \right\rbrace 
\end{eqnarray*}
be the simplex of probability $ M_{0} $-vectors.
One can define the parametric family of models as follows

\begin{eqnarray*}
\mathcal{F}=\left\lbrace  F_{\theta}= \left( f_{1}(.,\theta),...,f_{M_{0}}(., \theta)\right)^{T}: \theta \in \Theta \right\rbrace, 
\end{eqnarray*}
where  $ \Theta $ is a compact subset of $ k$- dimensional Euclidean space ($ k< M_{0}-1,~ \forall M_{0} \geq 1 $). 

 We assume that the probability distributions $F_{\theta}  $ is absolutely continuous with respect to a $  \sigma$-finite measure $ \mu $ on $ (\mathcal{X},\beta_{\mathcal{X}}) $. For simplicity $\mu  $ is either the Lebesgue measure or a counting measure. The parametric family of models may or may not contain the true model. If $ \mathcal{F} $ contains the true model, then there exists a  $ \theta_{0}\in \Theta $ such that $ F_{\theta_{0}}=F $ and the model $ F_{\theta} $ is said to be correctly specified. We are interested in testing 

\begin{eqnarray}\label{cond}
 H_{0}: F=F_{\theta_{0}} ~ \textsl{ versus}~ H_{1}: F\neq F_{\theta_{0}}. 
\end{eqnarray}

Note that  $ F(x)=(f_{1}(x),..., f_{M_{0}}(x))^{T}$ can be estimated by a bias reduced kernel estimator based on a random sample of size $n$; $X_{1} , ..., X_{M_{0}} $. In this following section, we present the brief review of this estimator.

\section{A Bias Reduced   Kernel Estimator}

Kernel density estimator was first introduced by Rosenblatt \cite{30} and Parzen \cite{31}. Suppose that $X_{1},..., X_{n}$ is a simple random sample from the unknown density function $f$. Let $K$ be a function on real line, i.e. the "kernel", and let $h$ be a positive value, i.e. the "bandwidth". Then the kernel density estimator of $f$ is defined as
\begin{eqnarray}\label{fn}
\hat{f}_{n,h}(x)=\frac{1}{nh}\sum_{i=1}^{n}K\left(\frac{x-X_{i}}{h} \right). 
\end{eqnarray}
To make the estimator meaningful, we introduce a measurable function $ K(.) $ that satisfies the following conditions.\\
(K.1) $ K(.) $ is of bounded variation on $ \mathbb{R}  $\\
(K.2) $ K(.) $ is right continuous on $ \mathbb{R}  $\\
(K.3) $ \parallel K \parallel_{\infty}=\sup_{x\in \mathbb{R}}\mid K(x)\mid<\infty  $\\
(K.4) $ \int_{\mathbb{R}}K(t)dt=1,$\\

Under the regularity conditions on $ K(.) $, let $f$ be  twice continuously differentiable in a neighborhood of $x$, then
\begin{eqnarray}\label{Bias}
Bias(\hat{f}_{n,h}(x)) &=&E(\hat{f}_{n,h}(x))-f(x)=\frac{h^{2}}{2}f^{''}(x)\int u^{2}K(u)du + o(h^{2})
\end{eqnarray}
and 
\begin{eqnarray}\label{Var}
Var(\hat{f}_{n,h}(x))=\frac{1}{nh}f(x)\int K^{2}(u)du +o((nh)^{-1})
\end{eqnarray}
Then from (\ref{Bias}) and (\ref{Var}) we have
\begin{eqnarray*}
MSE(\hat{f}_{n,h}(x))&=&Bias^{2}(\hat{f}_{n,h}(x))+Var(\hat{f}_{n,h}(x)) \\
             &=& \frac{1}{4}(f^{''}(x))^{2}h^{4}\left[ \int u^{2}K(u)du\right]^{2}+\frac{1}{nh}f(x)\int K^{2}(u)du+ o(h^{4}+(nh)^{-1} )          
\end{eqnarray*}
Devroy \cite{33} showed that the optimized bandwidth is $ h\sim O\left( n^{-\frac{1}{5}}\right) $ and then the optimal MSE is of the order $ n^{-\frac{4}{5}}.$
X.Xie, J.Wu \cite{29} introduced a new type of density estimator in order to reduce bias, investigated and calculated its bias, variance and MSE which show some improvement over the ordinary kernel density estimator.
Since the leading term of the bias is unavailable due to the unknown $f$, we can simply use its estimation to reduce the bias of the ordinary kernel density estimator, i.e. 
\begin{eqnarray*}
\hat{f}_{n,h}^{b}(x)=\hat{f}_{n,h}(x)-\widehat{Bias}(\hat{f}_{n,h}(x))
\end{eqnarray*}
As result, the proposed estimator is 
\begin{eqnarray}\label{fn1}
\hat{f}_{n,h}^{b}(x) &=&\hat{f}_{n,h}(x)-\frac{h^{2}}{2}\hat{f}_{n,h}^{''}(x) \int u^{2}K(u)du \\ \nonumber
                 &=&\frac{1}{nh}\sum_{i=1}^{n}K\left( \frac{x-X_{i}}{h}\right)-\frac{1}{2nh}\sum_{i=1}^{n}K^{''}\left( \frac{x-X_{i}}{h}\right) \int u^{2}K(u)du ,
\end{eqnarray}
where  $ \hat{f}_{n,h}(x) $ is given by (\ref{fn}).
From the way of construction, this new estimator should be able to reduce the bias and thus the MSE. To see whether this is the case or not, they next calculated the bias and the variance of  $ \hat{f}_{n,h}^{b}. $
These following regularity conditions on $f$ , $K$ and $h$ are in need:\\
\begin{enumerate}
\item $\int uK(u)=0$ 

\item $f$ is fourth differentiable in a neighbourhood of $ x $

\item $ h\rightarrow 0 $ and $ nh\rightarrow \infty $ as $ n\rightarrow \infty $
\end{enumerate}

\begin{thm} \textbf{(X.Xie, J.Wu \cite{29})}. Under 1), 2) and 3), 
\begin{eqnarray}\label{6}
Bias(\hat{f}_{n,h}^{b}(x))=-\frac{h^{3}}{6}f^{'''}(x)\int u^{3}K(u)du+o(h^{3})
\end{eqnarray}
and 
\begin{eqnarray}\label{7}
Var(\hat{f}_{n,h}^{b}(x))=\frac{1}{2nh}f(x)\left( \int u^{2}K(u)du\right)^{2}\int K^{''}(u)du+ o((nh)^{-1}).
\end{eqnarray}
Consequently,
\begin{eqnarray*}
MSE(\hat{f}_{n,h}^{b}(x))=o(h^{6}+(nh)^{-1})
\end{eqnarray*}
Under the regularity conditions on $f$ , $K$ and $h$,  the optimal MSE is of the order $ O\left( n^{-\frac{6}{7}}\right)  $ with $ h=O\left( n^{-\frac{1}{7}}\right).  $
\end{thm}

\section{A Breif Review of  Bregman Divergence}

Divergences are distance-like functions, widely used to assess the similarity between two objects.
Several authors proposed generalized divergences which encompass these classical divergences:
\begin{enumerate}
\item Csiszar's divergence \cite{HAL4}, which is a generalization of Amari's $ \alpha $-divergence \cite{HAL5}. Both these divergences encompass the Kullback-Leibler(KL) divergence  and its dual.
\item  Bregman divergence \cite{HAL3,HAL6}, which encompasses the euclidean(EUC) distance, the KL divergence and the Itakura Saito(IS) divergence.
\end{enumerate}
As a distance, a divergence should be non-negative and separable. However, a divergence does not necessarily satisfy the triangle inequality and the symmetry axiom of a distance.\\
Bregman ( see \cite{21,22,23,24}) introduced for  a convex subset of a Hilbert space $\mathbf{S}$ and  $ \phi: \mathbf{S} \rightarrow   \mathbb{R} $ a continuously differentiable strictly convex
function; the Bregman divergence   $ D_{\phi}^{B}: \mathbf{S} \times \mathbf{S} \longrightarrow \mathbb{R}_{+} $ as follows
\begin{eqnarray}
D_{\phi}^{B}(p,q)=\phi(p)-\phi(q)-\langle p-q,\bigtriangledown \phi(q) \rangle,~\forall (p,q)\in \mathbf{S}^{2}
\end{eqnarray}
where $ \bigtriangledown \phi(y) $ stands for the gradient of $ \phi $ evaluated at $y$ and $ \langle .,.\rangle $ is the standard Hermitian dot product. Thus the Bregman divergences between two probability density functions $ f $ and $ g $ is given by 
\begin{eqnarray}\label{bregDef}
D_{\phi}^{B}(f(x),g(x)):=\int_{\mathbf{X}}\left( \phi(f(x))-\phi(g(x))-(f(x)-g(x))\phi^{'}(g(x))\right)dx
\end{eqnarray}
where $\mathbf{X} $ is a support of the two density functions, $ f $ and $ g $; and $ \phi^{'}(.) $ the derivative of $ \phi(.): \mathbb{R} \rightarrow \mathbb{R} $ respected to $ x $.
On the other hand Basu al. \cite{AT6}, Minami, Eguchi \cite{AT15} introduced the basic Beta-divergence  and many researchers investigated their applications including \cite{AT33,AT8}. The main motivation was to develop the link between Beta-divergence and Bregman divergence.\\
It is also interesting to note that, the Beta-divergence has to be defined in limiting case for $ \beta\rightarrow 0 $ as the Itakura-Saito distance and for $\beta \rightarrow 1 $ as the KL-divergence. For $ \beta\rightarrow 2 $, we obtain the standard squared Euclidean ($L_{2}$-norm) distance. 
Therefore one can check that the Beta-divergence can be generated from the Bregman divergence using the following strictly convex continuous function \cite{AT33}
\begin{eqnarray}\label{57}
\phi(t)=\left\{\begin{array}{lc}
\frac{c_{1}}{\beta(\beta-1)}t^{\beta}+c_{2}t+c_{3},~~\beta\neq 0,1\\ 
c_{1}t \log(t)+c_{2}t+c_{3},~~\beta=1\\
-c_{1}\log(t)+c_{2}t+c_{3},~~\beta=0 
\end{array}\right.
\end{eqnarray}
Here $ c_{1}, c_{2} $ and $ c_{3} $ are some constants.\\

\begin{thm} 
\textbf{(Liese and Vajda \cite{32})} If the Bregman divergence $ D_{\phi}^{B}(p,q)$ satisfies the homogeneity condition\footnote{ $
D_{\phi}^{B}(kp,kq)=k^{\beta}D_{\phi}^{B}(p,q)$, ~\textsf{for all}~ $ p,q,k > 0 $ \textsf{with} $ \beta $ \textsf{equal to} $ 2, 1, 0 $. 
 }  with 
$
\beta=
\left\{\begin{array}{lc}
 2\\
1\\
0\\
\end{array}\right.
$, then $ 
\phi(t)= 
\left\{\begin{array}{lc}
t^{2},\\
t \log(t),\\
-\log(t)
\end{array}\right.
$
the statements hold modulo affine functions. 
\end{thm}
Assuming  that the density $f$ is unknown and the density $ f_{\theta} $  is theoretically known and satisfies: $ \int_{\mathbb{R}^{M_{0}}}f_{\theta}(x)dx$ is finite, we estimate $ D_{\phi}^{B}(f(x),f_{\theta}(x)) $ by 
\begin{eqnarray}\label{D}
\hat{D}_{\phi}^{B}(\hat{f}_{n,h}^{b}(x),f_{\theta}(x))=\int_{A_{n}} \phi(\hat{f}_{n,h}^{b}(x))-\phi(f_{\theta}(x))-(\hat{f}_{n,h}^{b}(x)-f_{\theta}(x))\phi^{'}(f_{\theta}(x))dx.
\end{eqnarray}
where $ A_{n}=\lbrace x\in \mathbb{R}^{M_{0}}, \hat{f}_{n,h}^{b}(x) \geq \gamma_{n}\rbrace $ and $ \gamma_{n}\rightarrow 0 $ is a sequence of positive constants. In this following section, we will use the methods developed in \cite{34} to establish convergence results for our  estimator $ \hat{D}_{\phi}^{B}(\hat{f}_{n,h}^{b}(x),f_{\theta}(x)) $.

\section{Consistancy of the Bregman divergence estimator}

For proving such consistency results, one usually writes the difference $ \hat{D}_{\phi}^{B}(\hat{f}_{n,h}^{b}(x),f_{\theta}(x))-D_{\phi}^{B}(f(x),f_{\theta}(x)) $ as the sum of a
probabilistic term
 $  \hat{D}_{\phi}^{B}(\hat{f}_{n,h}^{b}(x),f_{\theta}(x))-\mathbb{E}\hat{D}_{\phi}^{B}(\hat{f}_{n,h}^{b}(x),f_{\theta}(x)) $, and a deterministic term\\
  $ \mathbb{E}\hat{D}_{\phi}^{B}(\hat{f}_{n,h}^{b}(x),f_{\theta}(x))-D_{\phi}^{B}(\hat{f}_{n,h}^{b}(x),f_{\theta}(x)) $, the so-called bias. Troughout the remainder of this paper,\\
   $\mathbb{E} \hat{D}_{\phi}^{B}(\hat{f}_{n,h}^{b}(x),f_{\theta}(x)) $ is given by
\begin{eqnarray*}
\mathbb{E} \hat{D}_{\phi}^{B}(\hat{f}_{n,h}^{b}(x),f_{\theta}(x)):= \int_{A_{n}}\left[ \mathbb{E}\phi(\hat{f}_{n,h}^{b}(x))-\phi(f_{\theta}(x))-(\mathbb{E}\hat{f}_{n,h}^{b}(x)-f_{\theta}(x))\phi^{'}(f_{\theta}(x))\right] dx,
\end{eqnarray*}
 where $ A_{n} $ is defined in (\ref{D}).\\  
\begin{lem}  Let $ K(.) $ satisfy (K1)-(K4), let $ f(.) $ be a continuous bounded density, $ \phi $ be linear and strictly convex function and satisfies the Jensen inequality. Then, for each pair of sequence $ (a_{n})_{n\geq 1},~(b_{n})_{n\geq 1} $ such that $ 0< a_{n}< b_{n}\leq 1 $  with $ b_{n}\rightarrow 0 $ and $ na_{n}/\log(n)\rightarrow \infty $ as $ n\rightarrow \infty $, we have with probability $ 1 $
\begin{eqnarray*}
\sup_{a_{n}\leq h\leq b_{n}}\mid \hat{D}_{\phi}^{B}(\hat{f}_{n,h}^{b}(x),f_{\theta}(x))-\mathbb{E} \hat{D}_{\phi}^{B}(\hat{f}_{n,h}^{b}(x),f_{\theta}(x))\mid = 0\left( \sqrt{\frac{\log (1/a_{n})\vee \log~\log n}{n a_{n}}}\right).
\end{eqnarray*}
\end{lem}

\begin{pf}
 Define
\begin{eqnarray*}
\Delta_{n1}:=\hat{D}_{\phi}^{B}(\hat{f}_{n,h}^{b}(x),f_{\theta}(x))-\mathbb{E} \hat{D}_{\phi}^{B}(\hat{f}_{n,h}^{b}(x),f_{\theta}(x)).
\end{eqnarray*} 
We have 
\begin{eqnarray*}
\mid \Delta_{n1} \mid &=&\mid \int_{A_{n}}\left[\phi(\hat{f}_{n,h}^{b}(x))-\phi{(f_{\theta}(x))}-(\hat{f}_{n,h}^{b}(x)-f_{\theta}(x))\phi^{'}(f_{\theta}(x))\right] +\\
                      & & - \left[ \mathbb{E}\phi(\hat{f}_{n,h}^{b}(x))-\phi(f_{\theta}(x))-(\mathbb{E}\hat{f}_{n,h}^{b}(x)-f_{\theta}(x))\phi^{'}(f_{\theta}(x))\right]dx \vert\\
                      &\leq & \mid \int_{A_{n}}\left( \phi(\hat{f}_{n,h}^{b}(x))-\mathbb{E}\phi(\hat{f}_{n,h}^{b}(x))\right) dx\mid + \mid \int_{A_{n}}\left( \hat{f}_{n,h}^{b}(x)-\mathbb{E}\hat{f}_{n,h}^{b}(x)\right) \phi^{'}(f_{\theta}(x)) dx\mid \\
 \end{eqnarray*} 
 
  $ \phi $ verifies the Jensen inequality,  $ i.e.~ \mathbb{E}(\phi(\hat{f}_{n,h}^{b}(x)))\geq \phi(\mathbb{E}\hat{f}_{n,h}^{b}(x)) $ and  $ \phi $ is linear, $ i.e.~ \phi(\hat{f}_{n,h}^{b}(x))+ \phi(\mathbb{E}\hat{f}_{n,h}^{b}(x))=\phi (\hat{f}_{n,h}^{b}(x)+\mathbb{E}\hat{f}_{n,h}^{b}(x)) $. Therefore                                   
  \begin{eqnarray*}
\mid \Delta_{n1} \mid  \leq  \phi(\sup_{a_{n}\leq h \leq b_{n}}\mid \hat{f}_{n,h}^{b}(x))-\mathbb{E}\hat{f}_{n,h}^{b}(x) \mid)\int_{A_{n}} dx + \sup_{a_{n}\leq h \leq b_{n}} \mid \hat{f}_{n,h}^{b}(x))-\mathbb{E}\hat{f}_{n,h}^{b}(x) \mid \int_{A_{n}} \phi^{'}(f_{\theta}(x))dx.
\end{eqnarray*}
For $ 0< a_{n}< b_{n}\leq 1 $, we have
$\sup_{a_{n}\leq h \leq b_{n}} \mid f_{n,h}(x)-\mathbb{E}\hat{f}_{n,h}^{b}(x)\mid \leq \parallel \hat{f}_{n,h}^{b}(x)-\mathbb{E}\hat{f}_{n,h}^{b}(x) \parallel_{\infty} $ where $ \parallel. \parallel_{\infty} $ denotes, the suppremum norm, i.e, $ \parallel \psi \parallel_{\infty}:=\sup_{x\in \mathbb{R}}\mid \psi(x)\mid$. \\
Therefore,
\begin{eqnarray*}
\mid \Delta_{n1} \mid &\leq &   \phi\left( \parallel \hat{f}_{n,h}^{b}(x)-\mathbb{E}\hat{f}_{n,h}^{b}(x) \parallel_{\infty}\right)\int_{A_{n}}dx+  \parallel \hat{f}_{n,h}^{b}(x)-\mathbb{E}\hat{f}_{n,h}^{b}(x) \parallel_{\infty}\int_{A_{n}} \phi^{'}(f_{\theta}(x))dx.
\end{eqnarray*}

\begin{eqnarray*}
\sup_{a_{n}\leq h \leq b_{n}}\mid \Delta_{n1} \mid &\leq &   \phi\left( \sup_{a_{n}\leq h \leq b_{n}}\parallel \hat{f}_{n,h}^{b}(x)-\mathbb{E}\hat{f}_{n,h}^{b}(x) \parallel_{\infty}\right)\int_{A_{n}}dx+  \sup_{a_{n}\leq h \leq b_{n}} \parallel \hat{f}_{n,h}^{b}(x)-\mathbb{E}\hat{f}_{n,h}^{b}(x) \parallel_{\infty}\int_{A_{n}} \phi^{'}(f_{\theta}(x))dx.
\end{eqnarray*}
Finaly
\begin{eqnarray}\label{11}
\sup_{a_{n}\leq h \leq b_{n}}\mid \Delta_{n1} \mid &\leq &   \phi\left( \sup_{a_{n}\leq h \leq b_{n}}\parallel \hat{f}_{n,h}^{b}(x)-\mathbb{E}\hat{f}_{n,h}^{b}(x) \parallel_{\infty}\right)\int_{\mathbb{R}^{M_{0}}}dx+  \sup_{a_{n}\leq h \leq b_{n}} \parallel \hat{f}_{n,h}^{b}(x)-\mathbb{E}\hat{f}_{n,h}^{b}(x) \parallel_{\infty}\int_{\mathbb{R}^{M_{0}}} \phi^{'}(f_{\theta}(x))dx.
\end{eqnarray}
 Whenever $ K(.) $ is measurable and satisfies (K3)-(K4) and by the remark 2 in \cite{36}, when $ f(.)$ is bounded, for each pair of sequence $(a_{n})_{n\geq 1}$ and $(b_{n})_{n\geq 1}$ such that  $ 0< a_{n}< b_{n}\leq 1 $  with $ b_{n}\rightarrow 0 $ and $ na_{n}/\log(n)\rightarrow \infty $ as $ n\rightarrow \infty $, we have with probability $ 1 $
\begin{eqnarray}\label{12}
\sup_{a_{n}\leq h \leq b_{n}} \parallel \hat{f}_{n,h}^{b}(x)-\mathbb{E}\hat{f}_{n,h}^{b}(x) \parallel_{\infty}=0\left( \sqrt{\frac{\log (1/a_{n})\vee \log~\log n}{n a_{n}}}\right).
\end{eqnarray}
Since $  \int_{\mathbb{R}^{M_{0}}} dx <\infty $ and $ \int_{\mathbb{R}^{M_{0}}} \phi^{'}(f_{\theta}(x))dx< \infty $, in view of (\ref{11}) and (\ref{12}), we obtain with probability 1,
\begin{eqnarray*}
\sup_{a_{n}\leq h \leq b_{n}}\mid \Delta_{n1} \mid =0\left( \sqrt{\frac{\log (1/a_{n})\vee \log~\log n}{n a_{n}}}\right)+0\left( \sqrt{\frac{\log (1/a_{n})\vee \log~\log n}{n a_{n}}}\right).
\end{eqnarray*}
Thus 
\begin{eqnarray}
\sup_{a_{n}\leq h \leq b_{n}}\mid \Delta_{n1} \mid =0\left( \sqrt{\frac{\log (1/a_{n})\vee \log~\log n}{n a_{n}}}\right).
\end{eqnarray}
It concludes the proof of the lemma.
\end{pf}
\begin{lem} Let $ K(.) $ satisfy (K1)-(K4), let $ f(.) $ be a continuous bounded density, $ \phi $ linear and strictly convex function and satisfies the Jensen inequality. Then, for each pair of sequence $ (a_{n})_{n\geq 1},~(b_{n})_{n\geq 1} $ such that $ 0< a_{n}< b_{n}\leq 1 $   with $ b_{n}\rightarrow 0 $  as $ n\rightarrow \infty $,
\begin{eqnarray*}
\sup_{a_{n}\leq h\leq b_{n}}\mid \mathbb{E} \hat{D}_{\phi}^{B}(\hat{f}_{n,h}^{b}(x),f_{\theta}(x))-{D}_{\phi}^{B}(f(x),f_{\theta}(x))\mid = 0\left( b_{n}\right).
\end{eqnarray*}
\end{lem}

\begin{pf} Let $ \Delta_{n2}= \mathbb{E} \hat{D}_{\phi}^{B}(\hat{f}_{n,h}^{b}(x),f_{\theta}(x))-{D}_{\phi}^{B}(f(x),f_{\theta}(x)) $ ,
therefore
\begin{eqnarray*}
\mid \Delta_{n2} \mid &=&\mid\int_{A_{n}}\left[ \mathbb{E}\phi(\hat{f}_{n,h}^{b}(x))-\phi(f_{\theta}(x))-(\mathbb{E}\hat{f}_{n,h}^{b}(x)-f_{\theta}(x))\phi^{'}(f_{\theta}(x))\right] dx+\\
                      & & \int_{A_{n}}\left[ \phi(f(x))-\phi(f_{\theta}(x))-(f(x)-f_{\theta}(x))\phi^{'}(f_{\theta}(x))\right] dx\mid
\end{eqnarray*}
Repeat the arguments above in the terms $ \mid \Delta_{n1} \mid $ with the formal change of $ \hat{f}_{n,h}^{b} $ by $ f$. We show that, for any $ n\geq 1,$ 
\begin{eqnarray*}
\mid \Delta_{n2} \mid \leq \phi \left(  \sup_{a_{n}\leq h \leq b_{n}} \mid \hat{f}_{n,h}^{b}(x)-f(x)\mid\right) \int_{A_{n}}dx+ \sup_{a_{n}\leq h \leq b_{n}} \mid \hat{f}_{n,h}^{b}(x)-f(x)\mid \int_{A_{n}}\phi^{'}(f_{\theta}(x))dx,
\end{eqnarray*}
which implies
\begin{eqnarray}
\mid \Delta_{n2} \mid \leq \phi \left(  \sup_{a_{n}\leq h \leq b_{n}} \mid \hat{f}_{n,h}^{b}(x)-f(x)\mid\right) \int_{\mathbb{R}^{M_{0}}}dx+ \sup_{a_{n}\leq h \leq b_{n}} \mid \hat{f}_{n,h}^{b}(x)-f(x)\mid \int_{\mathbb{R}^{M_{0}}}\phi^{'}(f_{\theta}(x))dx.
\end{eqnarray}
In \cite{36}, when the density $f(.)$ is uniformly  continuous, we have for each pair of sequence $ (a_{n})_{n\geq 1},~(b_{n})_{n\geq 1} $ such that $ 0< a_{n}< b_{n}\leq 1 $, with $ b_{n}\longrightarrow 0 $ as $n \rightarrow \infty $,
\begin{eqnarray}\label{16}
\sup_{a_{n}\leq h \leq b_{n}} \parallel \hat{f}_{n,h}^{b}(x)-f(x) \parallel_{\infty}=0(b_{n}).
\end{eqnarray}
Thus,
\begin{eqnarray*}
\sup_{a_{n}\leq h \leq b_{n}} \mid \Delta_{n2} \mid \leq \phi \left(  \sup_{a_{n}\leq h \leq b_{n}} \parallel \hat{f}_{n,h}^{b}(x)-f(x)\parallel_{\infty}\right) \int_{\mathbb{R}^{M_{0}}}dx+ \sup_{a_{n}\leq h \leq b_{n}} \parallel \hat{f}_{n,h}^{b}(x)-f(x)\parallel_{\infty} \int_{\mathbb{R}^{M_{0}}}\phi^{'}(f_{\theta}(x))dx,
\end{eqnarray*}
where $ \int_{\mathbb{R}^{M_{0}}}dx $ and $  \int_{\mathbb{R}^{M_{0}}}\phi^{'}(f_{\theta}(x))dx $ are finite.
Then, in view of  (\ref{16})
\begin{eqnarray*}
\sup_{a_{n}\leq h \leq b_{n}} \mid \Delta_{n2} \mid =0(b_{n})+0(b_{n}).
\end{eqnarray*}
Finaly,
\begin{eqnarray}
\sup_{a_{n}\leq h \leq b_{n}} \mid \Delta_{n2} \mid =0(b_{n})
\end{eqnarray}
is deduced the proof of the lemma.
\end{pf}

\begin{thm}  Let $ K(.) $ satisfy (K3)-(K4), $ f(.) $ be a uniform, bounded and continuous density and $ \phi $ linear and strictly convex function and satisfy the Jensen inequality. Then, for each pair of sequence $ (a_{n})_{n\geq 1},~(b_{n})_{n\geq 1} $ such that $ 0< a_{n}< b_{n}\leq 1 $   with $ b_{n}\rightarrow 0 $ and $ na_{n}/\log(n)\rightarrow \infty $  as $ n\rightarrow \infty $, we have with probability $ 1 $
\begin{eqnarray*}
\sup_{a_{n}\leq h\leq b_{n}}\mid  \hat{D}_{\phi}^{B}(\hat{f}_{n,h}^{b}(x),f_{\theta}(x))-{D}_{\phi}^{B}(f(x),f_{\theta}(x))\mid = 0\left(\sqrt{\frac{\log (1/a_{n})\vee \log~\log n}{n a_{n}}} \vee b_{n}\right).
\end{eqnarray*}
\end{thm}

\begin{pf} We have 
\begin{eqnarray*}
\mid  \hat{D}_{\phi}^{B}(\hat{f}_{n,h}^{b}(x),f_{\theta}(x))-{D}_{\phi}^{B}(f(x),f_{\theta}(x))\mid &\leq & \mid \hat{D}_{\phi}^{B}(\hat{f}_{n,h}^{b}(x),f_{\theta}(x))-\mathbb{E} \hat{D}_{\phi}^{B}(\hat{f}_{n,h}^{b}(x),f_{\theta}(x))\mid  + \\
                        & & +\mid \mathbb{E} \hat{D}_{\phi}^{B}(\hat{f}_{n,h}^{b}(x),f_{\theta}(x))-{D}_{\phi}^{B}(f(x),f_{\theta}(x))\mid .
\end{eqnarray*}
Combinating the Lemma (1) and (2), we obtain 
\begin{eqnarray}
\sup_{a_{n}\leq h \leq b_{n}} \mid  \hat{D}_{\phi}^{B}(\hat{f}_{n,h}^{b}(x),f_{\theta}(x))-{D}_{\phi}^{B}(f(x),f_{\theta}(x))\mid =0\left( \sqrt{\frac{\log (1/a_{n})\vee \log~\log n}{n a_{n}}}\right)+ 0(b_{n}).
\end{eqnarray}
This entails that, as $ n\rightarrow \infty $,
\begin{eqnarray*}
\sup_{a_{n}\leq h \leq b_{n}} \mid  \hat{D}_{\phi}^{B}(\hat{f}_{n,h}^{b}(x),f_{\theta}(x))-{D}_{\phi}^{B}(f(x),f_{\theta}(x))\mid \longrightarrow 0.
\end{eqnarray*}
It concludes the proof of the Theorem.
\end{pf}

\section{Asymptotic behavior of Bregman divergence estimator}

Recall  the hypothesis testing (\ref{cond})  written as follows 
\begin{eqnarray*}
H_{0}: F_{\theta}= F \qquad \textsl{against}\qquad H_{1}:F_{\theta}\neq F. 
\end{eqnarray*} 
Note that for simplicity, we have omitted $ 0 $ on $ \theta $. We have to reject the null hypothesis iff $ D_{\phi}^{B}(\hat{F}_{n,h}^{b},F_{\hat{\theta}}) > d $ where $ d $ have to be chosen for getting a level $  \alpha $ test. In some situations it will be possible
to get the exact distribution of the statistic $D_{\phi}^{B}(\hat{F}_{n,h}^{b},F_{\hat{\theta}}) $ and then the value $d$. But in general this is not possible and we have to use the asymptotic distribution of
the statistic $D_{\phi}^{B}(\hat{F}_{n,h}^{b},F_{\hat{\theta}}) $. In this following theorem we present this asymptotic distribution.
\begin{thm} 
\label{thmAD1}
Let $ D_{\phi}^{B}(F,F_{\theta}) $ be the Bregman divergence  and let $ \hat{D}_{\phi}^{B}(\hat{F}_{n,h}^{b},F_{\hat{\theta}}) $ be its estimator . Under the null hypothesis  $ H_{0}: F_{\theta}=F $, we have 
\begin{eqnarray*}
2n\hat{D}_{\phi}^{B}(\hat{F}_{n,h}^{b},F_{\hat{\theta}}) \stackrel{\mathcal{L}}{\longrightarrow} \sum_{i=1}^{r}\beta_{i}Z_{i}^{2}+\sum_{j=1}^{s}\alpha_{j}Z_{j}^{2},
\end{eqnarray*}
when $ n\rightarrow \infty $. Where $ Z_{i}, i=1,...,r $ and $ Z_{j}, j=1,...,s $  are  iid normal variables with mean zero and variance $ 1$; we assume that $ r=s $.  $\beta_{i},i=1,...,r $ are the non null eigenvalues of the matrix $ H\Sigma_{F_{\theta}} $,  
$\alpha_{j},j=1,...,s $ are the non null eigenvalues of the matrix $ B\Sigma_{F} $, $ r=rank\left( \Sigma_{F_{\theta}}H\Sigma_{F_{\theta}} \right) $, and $ s=rank\left( \Sigma_{F}B\Sigma_{F} \right) $, being $ \Sigma_{F_{\theta}}=diag(F_{\theta})-F_{\theta}F_{\theta}^{t} $ and $ \Sigma_{F}=diag(F)-FF^{t} $
and
\begin{eqnarray*}
H=\left( \frac{\partial^{2}}{\partial f_{i} \partial f_{j}}{D}_{\phi}^{B}(F,F_{\theta})\right)_ {i,j=1,...,M_{0}},\qquad B=\left( \frac{\partial^{2}}{\partial f_{i\theta} \partial f_{j\theta}}{D}_{\phi}^{B}(F,F_{\theta})\right)_ {i,j=1,...,M_{0}}. 
\end{eqnarray*}
\end{thm}
\begin{pf} The second order Taylor  expansion of $ \hat{D}_{\phi}^{B}(\hat{F}_{n,h}^{b},F_{\hat{\theta}}) $ about $  F $ and $F_{\theta}$ gives
\begin{eqnarray*}
\hat{D}_{\phi}^{B}(\hat{F}_{n,h}^{b},F_{\hat{\theta}})= \frac{1}{2}(\hat{F}_{n,h}^{b}-F)^{T}B(\hat{F}_{n,h}^{b}-F)+\frac{1}{2}(F_{\hat{\theta}}-F_{\theta})^{T}H(F_{\hat{\theta}}-F_{\theta})+o(\parallel F_{n,h}-F \parallel^{2}+\parallel F_{\hat{\theta}}-F_{\theta} \parallel^{2} ).
\end{eqnarray*}
One can write
\begin{eqnarray*}
2 n \hat{D}_{\phi}^{B}(\hat{F}_{n,h}^{b},F_{\hat{\theta}})= \sqrt{n} (\hat{F}_{n,h}^{b}-F)^{T}B \sqrt{n} (\hat{F}_{n,h}^{b}-F)+ \sqrt{n} (F_{\hat{\theta}}-F_{\theta})^{T}H \sqrt{n} (F_{\hat{\theta}}-F_{\theta}) + 2n o(\parallel \hat{F}_{n,h}^{b}-F \parallel^{2}+\parallel F_{\hat{\theta}}-F_{\theta} \parallel^{2} ).
\end{eqnarray*}
And $ \sqrt{n} (F_{\hat{\theta}}-F_{\theta}) \stackrel{\mathcal{L}}{\longrightarrow}    N(0,\Sigma_{F_{\theta}} )$, when $ n\rightarrow \infty $; then $  \parallel F_{\hat{\theta}}-F_{\theta} \parallel^{2} = 0_{p}\left( n^{-1}\right)$. Therefore $ 2n o\left(\parallel F_{n,h}^{b}-F \parallel^{2} + \parallel F_{\hat{\theta}}-F_{\theta} \parallel^{2} \right) =o_{p}(1). $
The random variables $ 2 n \hat{D}_{\phi}^{B}(\hat{F}_{n,h}^{b},F_{\hat{\theta}}) $ and $\sqrt{n} (\hat{F}_{n,h}^{b}-F)^{T}B \sqrt{n} (\hat{F}_{n,h}^{b}-F)+ \sqrt{n} (F_{\hat{\theta}}-F_{\theta})^{T}H \sqrt{n} (F_{\hat{\theta}}-F_{\theta}) $ have the same asymptotic distribution.
Now by corollary 2.1 in Dik and Gunst \cite{39} the result follows.
\end{pf}

We consider now the case when the model is not specified \textit{i.e.} $ H_{1}:F_{\theta} \neq F   $. 
Let us  introduce the two important regularity assumptions.\\
-$ (A_{1}) $ Under the regularity conditions on the dominated model, the MLE is unique and asymptoticly normal under $ F_{\theta},~\forall \theta $
\begin{eqnarray*}
& & 1) \sqrt{n}(\hat{\theta}-\theta_{0})\Longrightarrow N(0, I(\theta_{0})^{-1})~~\textit{where}~~ I(\theta_{0})~is~ \textit{Fisher information}\\
& & 2)~F_{\hat{\theta}}\stackrel{as}{\longrightarrow} F_{\theta_{0}}~~\textit{when}~~n\rightarrow \infty.
\end{eqnarray*}
-$ (A_{2}) $ There exists  $ \theta\in \Theta; \wedge^{\ast} = \left(
\begin{array}{cc}
\wedge_{11} & \wedge_{12}\\
\wedge_{21} & \wedge_{22}
\end{array}
\right)
$
,with $ \wedge_{12}=\wedge_{21} $ and   such that 
\begin{eqnarray*}
\sqrt{n}\left(
\begin{array}{cc}
\hat{F}_{n,h}^{b}-F\\
F_{\hat{\theta}}-F_{\theta}
\end{array}
\right)
\stackrel{\mathcal{L}}{\longrightarrow} N(0,\wedge^{\ast}).
\end{eqnarray*}

\begin{thm}
\label{thm5}
 Under $ H_{1}:F_{\theta}\neq F  $ and we assume that the conditions $ (A_{1}), (A_{2}) $  hold, we have:
\begin{eqnarray*}
\sqrt{n}\left[  \hat{D}_{\phi}^{B}(\hat{F}_{n,h}^{b},F_{\hat{\theta}})-{D}_{\phi}^{B}(F,F_{\theta}) \right]  \stackrel{\mathcal{L}}{\longrightarrow} N(0,\wedge_{\phi}^{2})
\end{eqnarray*}
where 
\begin{eqnarray}\label{varia}
\wedge_{\phi}^{2}=K^{T}\wedge_{11}K+ K^{T}\wedge_{12}N+ N^{T}\wedge_{12}K+N^{T}\wedge_{22}N
\end{eqnarray}
$ K^{T}=(k_{1},...,k_{M_{0}}) $ with
\begin{eqnarray*}
k_{i}=\left(\frac{\partial}{\partial f_{i}} {D}_{\phi}^{B}(F,F_{\theta})\right),\qquad i=1,...,M_{0}.
\end{eqnarray*}
$ N^{T}=(n_{1},...,n_{M_{0}}) $ with
\begin{eqnarray*}
n_{i}=\left(\frac{\partial}{\partial f_{i\theta}} {D}_{\phi}^{B}(F,F_{\theta})\right), \qquad i=1,...,M_{0}. 
\end{eqnarray*}
\end{thm}
\begin{pf} A first order Taylor expansion gives
\begin{eqnarray*}
\hat{D}_{\phi}^{B}(\hat{F}_{n,h}^{b},F_{\hat{\theta}})={D}_{\phi}^{B}(F,F_{\theta})+K^{T}(\hat{F}_{n,h}^{b}-F)+ N^{T}(F_{\hat{\theta}}-F_{\theta})+o(\parallel F_{n,h}^{b}-F\parallel+\parallel F_{\hat{\theta}}-F_{\theta}\parallel).
\end{eqnarray*}
One can write
\begin{eqnarray*}
\sqrt{n}\left[ \hat{D}_{\phi}^{B}(\hat{F}_{n,h}^{b},F_{\hat{\theta}})-{D}_{\phi}^{B}(F,F_{\theta})\right] =\sqrt{n}\left[  K^{T}(\hat{F}_{n,h}^{b}-F)+ N^{T}(F_{\hat{\theta}}-F_{\theta})\right] +\sqrt{n}  o(\parallel \hat{F}_{n,h}^{b}-F\parallel+\parallel F_{\hat{\theta}}-F_{\theta}\parallel).
\end{eqnarray*}
Since 
$ \sqrt{n} (F_{\hat{\theta}}-F_{\theta}) \stackrel{\mathcal{L}}{\longrightarrow}    N(0,\Sigma_{F_{\theta}} )$, when $ n\rightarrow \infty $, with $ \Sigma_{F_{\theta}} $ defined in theorem (\ref{thmAD1}); then $  \parallel F_{\hat{\theta}}-F_{\theta} \parallel = 0_{p}\left( n^{-1/2}\right)$ and  $ \sqrt{n}  o\parallel F_{\hat{\theta}}-F_{\theta} \parallel = o_{p}\left( 1\right)$. Therefore $ \sqrt{n} o\left(\parallel F_{n,h}^{b}-F \parallel + \parallel F_{\hat{\theta}}-F_{\theta} \parallel \right) =o_{p}(1). $\\
Hence
\begin{eqnarray*}
\sqrt{n}\left[ \hat{D}_{\phi}^{B}(\hat{F}_{n,h}^{b},F_{\hat{\theta}})-{D}_{\phi}^{B}(F,F_{\theta})\right] =\sqrt{n}\left[  K^{T}(\hat{F}_{n,h}^{b}-F)+ N^{T}(F_{\hat{\theta}}-F_{\theta})\right] +  o_{p}(1)
\end{eqnarray*}
The random variables $ \sqrt{n}\left[ \hat{D}_{\phi}^{B}(\hat{F}_{n,h}^{b},F_{\hat{\theta}})-{D}_{\phi}^{B}(F,F_{\theta})\right] $ and $ \sqrt{n}\left[  K^{T}(\hat{F}_{n,h}^{b}-F)+ N^{T}(F_{\hat{\theta}}-F_{\theta})\right] $ have the same asymptotic distribution. \\
In view of $ A_{1} $ and $ A_{2} $ we have 
\begin{eqnarray*} 
\sqrt{n}\left[  K^{T}(\hat{F}_{n,h}^{b}-F)+ N^{T}(F_{\hat{\theta}}-F_{\theta})\right]\stackrel{\mathcal{L}}{\longrightarrow} N(0,\wedge_{\phi}^{2})
\end{eqnarray*} 
where $ \wedge_{\phi}^{2} $ is given by (\ref{varia}).
This completes the proof.
\end{pf}

\begin{rmk}
On the basis of the theorem \ref{thm5}, the power function at $ F\neq F_{\theta}  $ when testing $ H_{0}: F=F_{\theta} $ is given by the formula 
\begin{eqnarray}\label{power1}
\beta_{n,\phi}(F)=1-\Phi_{n}\left(\frac{t_{\alpha}-2n D_{\phi}^{B}(F,F_{\theta})}{2\sqrt {n}\wedge_{\phi}} \right); 
\end{eqnarray}
for a sequance of distribution function $ \Phi_{n}(x) $ tending uniformily to the standard normal distribution function $ \Phi(x);~t_{\alpha}$ is the critical value of $ T_{\phi}=2n \hat{D}_{\phi}^{B}(\hat{F}_{n,h}^{b},F_{\hat{\theta}}) $ and $ \wedge_{\phi} $ is given in thereom \ref{thm5}.\\
\end{rmk}

Thus thanks to the goodness-of-fit test, it is possible to chose the best model among a collection of candidate models to be the one which is close to the true distribution according to the Bregman divergence.

\section{Model Selection and Bregman Divergence  Based Test Statistic}

Consider  the situation in which we  have  two candidate parametric models $ F_{\theta} $ and $ F_{\gamma}=\lbrace  F(., \gamma); \gamma \in  \Gamma \subseteq \mathbb{R}^{M_{0}} \rbrace $ another candidate model. We would like to chose the best of two candidate models based on their discrimination statistic between the observations and models $ F_{\theta} $ and $ F_{\gamma} $  defined respectively as follows
$\hat{D}_{\phi}^{B}(\hat{F}_{n,h}^{b},F_{\hat{\theta}})$ and $\hat{D}_{\phi}^{B}(\hat{F}_{n,h}^{b},F_{\hat{\gamma}})$\\
Our major work is to propose some tests for model selection, i.e. for the null hypothesis \\
$ H_{0}: D_{\phi}^{B}(F,F_{\theta})=D_{\phi}^{B}(F,F_{\gamma}) $ ~means that the two models are equivalent,\\
 $ H_{f_{\theta}}:D_{\phi}^{B}(F,F_{\theta}) < D_{\phi}^{B}(F,F_{\gamma})$ ~means that $ F_{\theta}$ is better than $ F_{\gamma} $, \\
 $ H_{f_{\gamma}}:D_{\phi}^{B}(F,F_{\theta}) > D_{\phi}^{B}(F,F_{\gamma})$ ~means that $ F_{\theta}$ is worse than $ F_{\gamma} $. \\  
  
 To define the  model selection statistic, let us give this next lemma

\begin{lem}\label{lemm3}
 Under the assumptions of theorem \ref{thm5}, we have\\
\textit{(i) for the model $ F_{\theta} $}
\begin{eqnarray}
\hat{D}_{\phi}^{B}(\hat{F}_{n,h}^{b},F_{\hat{\theta}})={D}_{\phi}^{B}(F,F_{\theta})+T_{\theta}^{T}(\hat{F}_{n,h}^{b}-F)+ V_{\theta}^{T}(F_{\hat{\theta}}-F_{\theta})+o_{p}(1).
\end{eqnarray}

\textit{(ii) for model $F_{\gamma}  $}
\begin{eqnarray}
\hat{D}_{\phi}^{B}(\hat{F}_{n,h}^{b},F_{\hat{\gamma}})={D}_{\phi}^{B}(F,F_{\gamma})+T_{\gamma}^{T}(\hat{F}_{n,h}^{b}-F)+ V_{\gamma}^{T}(F_{\hat{\gamma}}-F_{\gamma})+o_{p}(1).
\end{eqnarray}
with
$ T_{\theta}^{T}=(t_{1},...,t_{M_{0}}) $ where
\begin{eqnarray*}
t_{i}=\left(\frac{\partial}{\partial f_{i}} {D}_{\phi}^{B}(F,F_{\theta})\right),\qquad i=1,...,M_{0}
\end{eqnarray*}
and
$ V_{\theta}^{T}=(v_{1},...,v_{M_{0}}) $ with
\begin{eqnarray*}
v_{i}=\left(\frac{\partial}{\partial f_{i\theta}} {D}_{\phi}^{B}(F,F_{\theta})\right) ,\qquad i=1,...,M_{0}. 
\end{eqnarray*}

\end{lem}
\begin{pf} 
The results follows from a first order Taylor expansion.
\end{pf}

We define
\begin{eqnarray}\label{kappa}
\kappa^{2}=(T_{\theta}-T_{\gamma}; V_{\theta}-V_{\gamma})^{T}\wedge^{\ast}(T_{\theta}-T_{\gamma}; V_{\theta}-V_{\gamma})
\end{eqnarray}
which is the variance of 
\begin{eqnarray*}
\sqrt{n}(T_{\theta}-T_{\gamma}; V_{\theta}-V_{\gamma})^{T}\left(
\begin{array}{cc}
F_{n,h}-F\\
F_{\hat{\theta}}-F_{\theta}
\end{array}
\right)
\end{eqnarray*}

Since $ T_{\theta},T_{\gamma}, V_{\theta}, V_{\gamma} $ and $ \wedge^{\ast}, $ consistently estimated by  their sample analogues  $ T_{\hat{\theta}},T_{\hat{\gamma}}, V_{\hat{\theta}}, V_{\hat{\gamma}} $ and $ \hat{\wedge}^{\ast}. $
Hence $ \kappa^{2} $ is consistently estimated by 
\begin{eqnarray*}
\hat{\kappa}^{2}=(T_{\hat{\theta}}-T_{\hat{\gamma}}; V_{\hat{\theta}}-V_{\hat{\gamma}})^{T}\hat{\wedge}^{\ast}(T_{\hat{\theta}}-T_{\hat{\gamma}}; V_{\hat{\theta}}-V_{\hat{\gamma}})
\end{eqnarray*}

Let $ U $ be the model selection statistic and be given by
\begin{eqnarray}\label{U}
U=\frac{\sqrt{n}}{\hat{\kappa}}\left[ \hat{D}_{\phi}^{B}(\hat{F}_{n,h}^{b},F_{\hat{\theta}})-\hat{D}_{\phi}^{B}(\hat{F}_{n,h}^{b},F_{\hat{\gamma}}) \right]
\end{eqnarray}
 
\begin{thm}
\label{thm7}
 (Asymptotic distribution of the U-statistic).\\
 Under the assumptions of theorem \ref{thm5}, suppose that $ \kappa\neq 0 $, then under the null hypothesis $ H_{0},~U\longrightarrow N(0,1). $
\end{thm}

\begin{pf} 
  From the lemma \ref{lemm3}, It follows that 
\begin{eqnarray*}
\hat{D}_{\phi}^{B}(\hat{F}_{n,h}^{b},F_{\hat{\theta}})-\hat{D}_{\phi}^{B}(\hat{F}_{n,h}^{b},F_{\hat{\gamma}})&=&D_{\phi}^{B}(F,F_{\theta})-D_{\phi}^{B}(F,F_{\gamma})+T_{\theta}^{T}(\hat{F}_{n,h}^{b}-F)-T_{\gamma}^{T}(\hat{F}_{n,h}^{b}-F)+\\
& & + V_{\theta}^{T}(F_{\hat{\theta}}-F_{\theta})-V_{\gamma}^{T}(F_{\hat{\gamma}}-F_{\gamma})+o_{p}(1).
\end{eqnarray*}
Under $ H_{0},~ D_{\phi}^{B}(F,F_{\theta})=D_{\phi}^{B}(F,F_{\gamma}), ~F_{\theta}=F_{\gamma} $ and $ F_{\hat{\theta}}=F_{\hat{\gamma}} $ we have
\begin{eqnarray*}
\hat{D}_{\phi}^{B}(\hat{F}_{n,h}^{b},F_{\hat{\theta}})-\hat{D}_{\phi}^{B}(\hat{F}_{n,h}^{b},F_{\hat{\gamma}})&=&T_{\theta}^{T}(\hat{F}_{n,h}^{b}-F)-T_{\gamma}^{T}(\hat{F}_{n,h}^{b}-F)+V_{\theta}^{T}(F_{\hat{\theta}}-F_{\theta})-V_{\gamma}^{T}(F_{\hat{\gamma}}-F_{\gamma})+o_{p}(1)\\
&=& (T_{\theta}-T_{\gamma}, V_{\theta}-V_{\gamma})^{T}\left(
\begin{array}{cc}
\hat{F}_{n,h}^{b}-F\\
F_{\hat{\theta}}-F_{\theta}
\end{array}
\right)
+o_{p}(1)
\end{eqnarray*}
Finally, applying the Central Limit Theorem and assumptions $(A_{1})-(A_{2})$, we can now immediately obtain
$ U\longrightarrow N(0,1). $
It concludes the proof of the theorem  \ref{thm7}.\\
\end{pf}
Theorem \ref{thm7} is quite general and gives us a wide variety of asymptotic standard normal tests for model selection based on Bregman divergence type statistic.

\section{Example and Simulation Study}
\subsection{Example with real data}

We analyze a real life-data set in which a selection between Gamma and log-normal distributions is of a prime interest.\\ 
\textbf{Data set}: Suppose the following observations (as given by Lieblein and Zelen (1956) for
the lifetime) are used to test whether the data come from a Gamma or a Log-Normal. The
data given arose in tests on endurance of deep groove ball bearings. The data are number of million revolutions before failure for each of the lifetime tests and they are:
17.88, 28.92, 33.00, 41.52, 42.12, 45.60, 48.80, 51.84, 51.96, 54.12, 55.56, 67.80, 68.44,
68.64, 68.88, 84.12, 93.12, 98.64, 105.12, 105.84, 127.92, 128.04, 173.40. Here we consider Gamma model as the component of vector of densities $ F_{\theta} $ and log-normal as the component of $ F_{\gamma} $ defined respectively in section 7. Therefore, to analyze a skewed positive data set an experimenter might wish to select one of them.\\

A random variable $X$ is said to have a Gamma distribution, denoted by
$ GA(\alpha,\eta) $ , when it has the probability density function (PDF) of
\begin{eqnarray*}\label{GA}
f_{GA}(x; \alpha, \eta)=\left\{\begin{array}{lc}
\frac{\eta^{\alpha}}{\Gamma(\alpha)}x^{\alpha-1}e^{-\eta x}~~~~~~~~~~~x\geq 0,\alpha>0, \eta > 0\\
0 ~~~~~~~~~~~~~~~~~~~~~~~~~~~~~~~~~~~~~~~~~~~~x< 0,
\end{array}\right.
\end{eqnarray*}
where $ \Gamma(\alpha)=\int_{0}^{\infty}x^{\alpha-1}e^{-x} dx $. We know that $ E(X)=\frac{1}{n}\sum_{i=1}^{n} X_{i}=\frac{\hat{\alpha}}{\hat{\eta}} $, then MLE of $ \eta $ in terms of $ \alpha $ is given by
\begin{eqnarray*}
\hat{\eta}_{n}=\frac{n\hat{\alpha}_{n}}{\sum_{i=1}^{n}X_{i}},
\end{eqnarray*}
The MLE $ \hat{\alpha}_{n} $ can not be easy to compute because its construction requires the
solution of the maximum likelihood equation (MLEq) 
\begin{eqnarray*}
\sum_{j=1}^{n}ln(\hat{\eta_{n}} X_{j})-\frac{n\dot{\Gamma}(\hat{\alpha}_{n})}{\Gamma(\hat{\alpha}_{n})}=0~~\textsl{and}~~  \sqrt{n}\left( \hat{\alpha}_{n}-\alpha \right)  \stackrel{\mathcal{L}}{\longrightarrow} N\left(0,I(\alpha)^{-1} \right),
\end{eqnarray*}
 when  $ n \rightarrow \infty$. The sequel dot means derivation w.r.t. $ \alpha $. We introduce the multi-step MLE-process \cite{MLE}, which in this case provides us an estimator $ \alpha^{\star}_{n} $ such that
$ \sqrt{n}\left( \alpha_{n}^{\star}-\alpha \right)  \stackrel{\mathcal{L}}{\longrightarrow} N\left(0,I(\alpha)^{-1} \right) $ when  $ n \rightarrow \infty $. Suppose that we have $n$ i.i.d. r.v.'s $ X^{n}=\left( X_{1},...,X_{n}\right)  $ with smooth density function
$ f(x,\alpha) $ and $ l(x,\alpha)=ln f(x,\alpha) $. Here $ \alpha \in \Theta $. Let us denote $ \bar{\alpha}_{N} $ the premilinary estimator constructed by the first $ N=\left[  n^{\delta} \right]  $ observations $ X^{N}=\left( X_{1},...,X_{N} \right)  $ with $ \delta = \left(\frac{1}{2}, 1 \right).$ Then the one-step MLE-process $ \alpha^{\star}_{n}=\left(\alpha^{\star}_{k,n},~N+1\leq k\leq n \right) $ is defined by the equality 
\begin{eqnarray*}
\alpha^{\star}_{k,n}=\bar{\alpha}_{N}+I\left( \bar{\alpha}_{N} \right)^{-1}\frac{1}{k}\sum_{j=N+1}^{k}\dot{l}\left(X_{j},\bar{\alpha}_{N}\right), 
\end{eqnarray*}
for $ k=\left[sn \right],~~s\in \left( 0,1\right]$; we have the convergence 
\begin{eqnarray*}
\sqrt{k}\left( \alpha^{\star}_{k,n}-\alpha_{0} \right)  \stackrel{\mathcal{L}}{\longrightarrow} N\left(0,I(\alpha_{0})^{-1} \right), 
\end{eqnarray*}
Here $  s $ is fixed and $ n \rightarrow \infty $. Therefore $ \alpha^{\star}_{n} $ \textit{is a good
estimator-process, i.e.,} $ \alpha^{\star}_{k,n} $  depends on $ X^{k}=\left(X_{1},...,X_{k} \right), $  easy to calculate and is asymptotically efficient because it is asymptotically equivalent to the MLE.\\
Therefore for our case, the preliminary estimator 
$\bar{\alpha}=\frac{\hat{\eta}}{n}\sum_{i=1}^{n}X_{i}\longrightarrow \alpha_{0},~\sqrt{n}(\bar{\alpha}-\alpha_{0})   \stackrel{\mathcal{L}}{\longrightarrow} N(0,I(\alpha_{0}^{-1}))$.
Then the one-step MLE-process is given by
\begin{eqnarray*}
\hat{\alpha}=\bar{\alpha}+\frac{1}{nI(\bar{\alpha})}\sum_{i=1}^{n}\left[ ln(\hat{\eta} X_{i})-\frac{\dot{\Gamma}(\bar{\alpha})}{\Gamma(\bar{\alpha})}\right].~\textsl{Then} \sqrt{n}(\hat{\alpha}-\alpha)\stackrel{\mathcal{L}}{\longrightarrow} N(0,I(\alpha)^{-1}),~n \rightarrow \infty
\end{eqnarray*}
 where  $ I(\bar{\alpha})=\frac{\left( \ddot{\Gamma}(\bar{\alpha})\Gamma(\bar{\alpha})-\dot{\Gamma}^{2}(\bar{\alpha}) \right)}{\Gamma^{2}(x)}$ with $ \dot{\Gamma}(\bar{\alpha})=\int_{0}^{\infty}(ln x) x^{\bar{\alpha}-1}e^{-x}dx,\ddot{\Gamma}(\bar{\alpha})=\int_{0}^{\infty}(ln x)^{2}x^{\bar{\alpha}-1}e^{-x}dx$
and $ \alpha_{0} $ the true value of $ \alpha $.

A random variable $X$ is distributed as Log-Normal, denoted as $ LN(\mu,\sigma^{2}) $, if $ ln(X) $ is normal , \textit{i.e.} $ ln(X)\sim N(\mu,\sigma^{2}) $. The probability density of $ X $ is given by 
\begin{eqnarray*}\label{LN}
f_{LN}(x; \mu, \sigma^{2})=\left\{\begin{array}{lc}
\frac{1}{\sqrt{2\pi}\sigma x}e^{-\frac{1}{2\sigma^{2}}(ln(x)-\mu)^{2}}~~~~~~~~~~~x\geq 0,\mu>0, \sigma > 0\\
0 ~~~~~~~~~~~~~~~~~~~~~~~~~~~~~~~~~~~~~~~~~~~~x< 0.
\end{array}\right.
\end{eqnarray*}
The MLE of $ \mu $ and $ \sigma $ are given below respectively:

\begin{eqnarray*}
\hat{\mu}=\frac{1}{n}\sum_{i=1}^{n}ln(X_{i})~\textit{and}~\hat{\sigma}^{2}=\frac{1}{n}\sum_{i=1}^{n}(ln(X_{i})-\hat{\mu})^{2}.
\end{eqnarray*}
For $ \beta=3 $ and $ c_{1}=1 $ the relations  (\ref{bregDef}) and (\ref{57})  allow us to compute  the $ \hat{D}_{\phi}^{B}(\hat{f}_{n,h}^{b}(x),f_{ \widehat{GA}}(x))$ and $ \hat{D}_{\phi}^{B}(\hat{f}_{n,h}^{b}(x),f_{ \widehat{LN}}(x)) $  as follows
\begin{eqnarray}\label{bre1}
\hat{D}_{\phi}^{B}(\hat{f}_{n,h}^{b}(x),f_{ \widehat{GA}}(x))=\frac{1}{6}\int_{0}^{\infty}\left( (\hat{f}_{n,h}^{b}(x))^{3}-3f_{n,h}^{b}(x)f_{\widehat{GA}}^{2}(x)+2f^{3}_{\widehat{GA}}(x)\right)dx
\end{eqnarray}
and
\begin{eqnarray}\label{bre2}
\hat{D}_{\phi}^{B}(\hat{f}_{n,h}^{b}(x),f_{ \widehat{LN}}(x))=\frac{1}{6}\int_{0}^{\infty}\left( (\hat{f}_{n,h}^{b}(x))^{3}-3f_{n,h}^{b}(x)f_{\widehat{LN}}^{2}(x)+2f^{3}_{\widehat{LN}}(x)\right)dx
\end{eqnarray}
where $ \hat{f}_{n,h}^{b}(.) $ is given by (\ref{fn1}). We consider  Gaussian kernel $ K(u)=\frac{1}{\sqrt{2\pi}}e^{-\frac{1}{2}u^{2}} $  because  it has infinitely many (nonzero) derivatives as our candidate models. Note that for the Gaussian kernel,
\begin{eqnarray*} 
\hat{f}_{n,h}^{b}(x)= \frac{1}{2\sqrt{2\pi} nh}\sum_{i=1}^{n}\left[3- \left( \frac{x-X_{i}}{h}\right)^{2}\right]e^{-\frac{1}{2}\left( \frac{x-X_{i}}{h}\right)^{2}}.
\end{eqnarray*}

 To get $ h $ optimal, the cross-validation method introduced by Rudemo (1982) and Bowman (1984) giving the simple and attractive smoothing parameter is used. Hence $ h\equiv 
h_{CV}=arg \min_{h>0}CV(h) $ where CV(h) is cross-validation given by $ CV(h)=\int \hat{f}^{2}dx-\frac{2}{n}\sum_{i=1}^{n}\hat{f},_{-i}(X_{i}) $ and $\hat{f},_{-i}(x)=\frac{1}{(n-1)h}\sum_{j\neq i} K\left( \frac{x-X_{j}}{h}\right)$.  $ f_{ \widehat{GA}}(x) $ and $ f_{\widehat{LN}}(x) $ are parametric estimators of Gamma and log-normal models and are given by
$
f_{\widehat{GA}}(x)=\frac{\hat{\eta}^{\hat{\alpha}}}{\Gamma(\hat{\alpha)}}x^{\hat{\alpha}-1}e^{-\hat{\eta} x}
$
and 
$
f_{ \widehat{LN}}(x)= \frac{1}{\sqrt{2\pi}\hat{\sigma} x}e^{-\frac{1}{2\hat{\sigma}^{2}}(ln(x)-\hat{\mu})^{2}}
$ respectively.

For the data at hand, Ali-Akbar Bromideh and Reza Valizadeh (2013)  proved that Gamma fits better in discrimination between Gamma and log-normal distributions using the Ratio of Minimized Kullback-Leibler Divergence. Therefore we obtain for the Gamma model $ \hat{\alpha}=4.028040  $ and $ \hat{\eta}=0.055767.$ And for log-normal distribution  $ \hat{\mu}=4.150614  $ and $ \hat{\sigma}=0.521485.$ 
From (\ref{bre1}) and (\ref{bre2}) one has $ \hat{D}_{\phi_{1}}^{B} \equiv \hat{D}_{\phi}^{B}(\hat{f}_{n,h}^{b}(x),f_{ \widehat{GA}}(x))=0.000021 $ and $ \hat{D}_{\phi_{2}}^{B}\equiv \hat{D}_{\phi}^{B}(\hat{f}_{n,h}^{b}(x),f_{ \widehat{LN}}(x))=0.000024 $.  Bregman divergence is the non-symmetric measure of the difference (dissimilarity) between two probability distributions. Being  interested to select the model minimizing the BD as the best model,  $ i.e~U=-0.000002  $. At 5\% significance level, we compare $U$ with $−1.96$ and $1.96$.  $U$ falls between $−1.96$ and $1.96$, we conclude that both estimated models fit the data equally well. The Figure $1$ and $2$ show that these models may provide similar data fit for moderate sample sizes. Note that many observed data are concentrated between $40$ and $80$ considering the axis of the $ X $ of our figures.

\begin{figure}[!h]
\begin{center}
\includegraphics[scale=0.7]{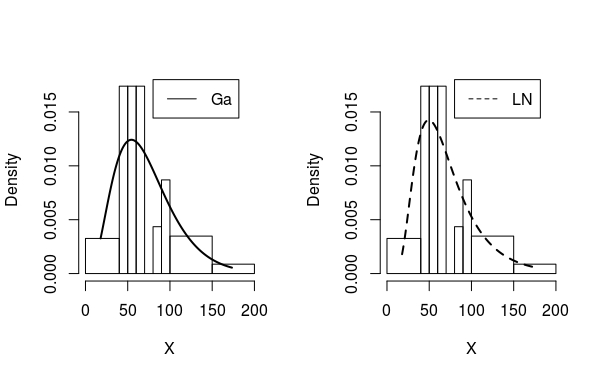}
\caption{The histogram  with  Log-Normale and Gamma density functions for the given data set.}
\end{center}
\end{figure}

\begin{figure}[!h]
\begin{center}
\includegraphics[scale=0.7]{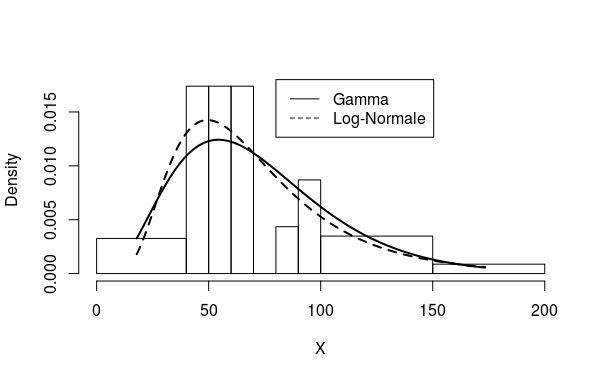}
\caption{The histogram  and two fitted density functions for the given data set.}
\end{center}
\end{figure}
\newpage
\subsection{Simulation study}

To illustrate well our model selection procedure,  we have defined our candidate models, Bregman divergence type statistic to measure the closness between our candidate models and reduced bias kernel density estimator. Consider the data generated from a mixture of Gamma and log-normal distributions. These two distributions are calibrated by the multi-step MLE process from the data set defined in sub-section $8.1$. 
Hence the Data Generating Process (DGP) has density
\begin{eqnarray*}
l(\pi)=\pi \textsl{Gamma}(4.02804, 0.05576722 ) + (1-\pi)\textsl{log-normal} (4.150614, 0.5214847 )
\end{eqnarray*}
where $ \pi\in (0,1) $ is specific to each set of experiments. In each set, several random samples are drawn from this mixture. The sample size varies from $20$ to $90$, and for each sample size the number of replications is $1000$. 
 We choose different values of $ \pi $ which are $ 0.00,0.25, 0.5, 0.75, 1.00. $  Although our proposed model selection procedure does not require that the data generating process belong
to either of the candidate models. We consider the two limiting cases  $\pi= 0.00  $ and $= 1.00$ for they correspond to the correctly specified cases. For $ \pi=0.25 $ and $ \pi=0.75 $ both candidate models are misspecified but not at equal distance from the DGP. These cases correspond
to a DGP which is Gamma or log-normal distributions but slightly contaminated by the other distribution. The value $ \pi=0.5 $ is the value for which the Gamma and log-normal distributions are approximately
at equal distance to the mixture $ l(\pi) $ according to statistics $ \hat{D}_{\phi_{1}}^{B} \equiv \hat{D}_{\phi}^{B}(\hat{f}_{n,h}^{b}(x),f_{ \widehat{GA}}(x)) $ and $ \hat{D}_{\phi_{2}}^{B} \equiv \hat{D}_{\phi}^{B}(\hat{f}_{n,h}^{b}(x),f_{ \widehat{LN}}(x)) $. Our model selection statistic is given by $ U$. The results of our five sets of experiments are presented in \textbf{Tables 1-5}. For $ n=60 $, we plot the histogram of datasets and overlay the curves for Gamma and log-normal distribution in order to analyse the closness of these two models.

\vspace*{0.2cm}
 
\textbf{Table 1. DGP}= $ \textsl{log-normal} (4.150614, 0.5214847 ) $\\

\begin{tabular}{lcccccr}
\hline
   n   & & 20 & 40  &   60   &   80  &  90   \\
 \hline
$ \hat{\alpha} $ & & 4.5391 &  4.1526   &  4.0529     &  3.9728  & 3.9540\\
                 & & (1.6452) &  (0.9603)  & (0.7491)  & (0.6603) &  (0.5743)         \\
                 &   &           &          &          &             &          \\
$ \hat{\eta} $  & &  0.0644 & 0.0579    &  0.0562   &  0.0551 &  0.0548      \\
                & & (0.2783) &  (0.0162)   & (0.0125)&  (0.0109) &   ( 0.0095)               \\
                 &   &           &          &          &             &          \\
$ \hat{\mu} $    & & 4.1494 &  4.1522   &  4.1515    & 4.1480 &   4.1483\\
                 & & (0.1155) &  (0.0859)   & (0.0686) & (0.0582)&  ( 0.0540)             \\
                  &   &          &          &          &             &          \\
$ \hat{\sigma} $  & &  0.5019 &  0.5121   &   0.5145    &  0.5180 &   0.5184\\
                  & & (0.0840) & ( 0.0566)  &   (0.0469)      & (0.0418)    & ( 0.0256)    \\
                   &   &           &          &          &             &          \\
$ \hat{D}_{\phi_{1}}^{B} $ & & 0.000031 &   0.000028 & 0.000026 & 0.000026 & 0.000026 \\
                           & & (0.000015)&  (0.000009) &  (0.000007)     & (0.000006) & (0.000006)    \\
                            &   &           &          &          &                     \\
$ \hat{D}_{\phi_{2}}^{B} $ & & 0.000026   & 0.000023  & 0.000021 & 0.000021  & 0.000021\\
                          & & (0.000014)  &  (0.000009)  & (0.000007) & (0.000006) &  (0.000004)    \\
                            &   &           &          &          &             &          \\
$ U $ & & 1.470805 &  2.371773    & 3.035950 & 3.132926  &  3.283792    \\
     &  & (0.000003) &  (0.000002)    & (0.000002) &  (0.000002)   & (0.000001)               \\                     
Model selection & Correct       & 24.9\%  & 67.0\%   & 87.1\%  & 91.7\% & 93.9\% \\
based on $ U $  & Indecisive  & 74.6\% & 32.8\% & 12.8\%  & 8.2\%  &   6.1\%\\
                & incorrect   &  0.5\% &  0.2\% & 0.1\%   & 0.1\%  & 0.0\% \\                 
 \hline             
\end{tabular}

\newpage

\textbf{Table 2. DGP}= $  \textsl{Gamma}(4.02804, 0.05576722 )  $

\vspace*{0.1cm}

\begin{tabular}{lcccccr}
\hline
   n             &  & 20      & 40  &   60   &   80  &  90   \\
 \hline
$ \hat{\alpha} $ & & 4.7264 &   4.3407    &   4.1839     & 4.2493   & 4.1606\\
                 & & (1.7632) &  (0.9604)  & (0.7605)  & (0.7903) &  (0.6042)         \\
                 &   &           &          &          &             &          \\
$ \hat{\eta} $  & &  0.6627 & 0.0606   &  0.0580   &  0.0589 &  0.0576\\
                & & (0.0262) &  (0.0148)   & (0.0187)&  (0.0113) &   ( 0.010)               \\
                 &   &           &          &          &             &          \\
$ \hat{\mu} $    & & 4.1499 &  4.1500  &  4.1508   & 4.1527 &   4.1525\\
                 & & (0.120811) &  (0.083108)   & (0.052709) & (0.068731)&  ( 0.0575)             \\
                  &   &           &          &          &             &          \\
$ \hat{\sigma} $  & &  0.509398 &  0.518293   &   0.526293    &  0.521632 &   0.5250\\
                  & & (0.095431) & ( 0.063427)  &   (0.034347)& (0.0 53744)    & ( 0.0432)    \\
                   &   &           &          &          &             &          \\
$ \hat{D}_{\phi_{1}}^{B} $ & & 0.000027 &   0.000024 & 0.000023 & 0.000023 & 0.000022 \\
                           & & (0.000012)&  (0.00008) &  (0.00009) & (0.000006) & (0.00000)    \\
                            &   &           &          &          &                     \\
$ \hat{D}_{\phi_{2}}^{B} $ & & 0.000030  & 0.000027  & 0.000026 & 0.000026  & 0.000025\\
                          & & (0.000012)  &  (0.00009)  & (0.000010) & (0.000006) &  (0.000005)    \\
                            &   &           &          &          &             &          \\
$ U $ & & -1.177846 &  -1.604843   & -1.968482 & -2.260217  &  -2.277816    \\
     &  & (0.000003) &  (0.000002)    & (0.000002) &  (0.000002)   & (0.000001)               \\           
Model selection & Correct         & 14.2\% & 31.6\% & 47.0\%  & 64.3\%  & 66.2\% \\
based on $ U $  & Indecisive   &  85.6\% & 68.0\%  & 52.9\%   &  35.4\% &  33.6\%      \\
                & Incorrect   &  0.2\%  & 0.4\%   & 0.1\%    & 0.3\%   & 0.2\%  \\
\hline             
\end{tabular}

\vspace*{0.2cm}

\textbf{Table 3. DGP}= $ 0.25 \textsl{Gamma}(4.02804, 0.05576722 ) + 0.75\textsl{log-normal} (4.150614, 0.5214847 ) $\\

\begin{tabular}{lcccccr}
\hline
   n             &  & 20      & 40  &   60   &   80  &  90   \\
 \hline
$ \hat{\alpha} $ & & 7.2498 &  6.7094    &   6.4145    & 6.3138   & 6.3687\\
                 & & (2.5978) &  (1.7032)  & (1.3363)  & (0.6603) &  (1.0345)         \\
                 &   &           &          &          &             &          \\
$ \hat{\eta} $  & &  0.101999 & 0.093505   &  0.088910   &  0.087376 &  0.08830\\
                & & (0.041302) &  (0.026984)   & (0.020856)&  (0.010946) &  ( 0.01631)      \\
                 &   &           &          &          &             &          \\
$ \hat{\mu} $    & & 4.200020 &  4.200743   &  4.200762    & 4.200372 &   4.1993\\
                 & & (0.091232) &  (0.065416)   & (0.051791) & (0.058290)&  ( 0.0422)             \\
                  &   &           &          &          &             &          \\
$ \hat{\sigma} $  & &  0.391877 &  0.398275   &   0.403963    &  0.404366 &   0.4030\\
                  & & (0.066546) & ( 0.046814)  &   (0.0 3.9236)      & (0.041885)    & ( 0.0314) \\
                   &   &           &          &          &             &          \\
$ \hat{D}_{\phi_{1}}^{B} $ & & 0.000041 &   0.000040 & 0.000039 & 0.000039 & 0.000036 \\
                           & & (0.000015)&  (0.000012) &  (0.000011) & (0.000009) & (0.000008)    \\
                            &   &           &          &          &                     \\
$ \hat{D}_{\phi_{2}}^{B} $ & & 0.000037  & 0.000036  & 0.000034 & 0.000034  & 0.000034\\
                          & & (0.000016)  &  (0.000012)  & (0.000011) & (0.000009) &  (0.00008)    \\
                            &   &           &          &          &             &          \\
$ U $ & & 0.815310 &  1.427280    & 2.106998 & 2.934055  &  3.455230    \\
     &  & (0.000005) &  (0.000003)    & (0.000003) &  (0.000002)   & (0.000002)               \\     
Model selection & Gamma & 1.9\% & 1.5\% & 0.6\%  & 0.3\% &  0.1\% \\
based on $ U $  & Indecisive &  93.9\% & 78.1\% & 43.5\%  & 10.5\% &  4.0\% \\
                & log-normal & 4.2\% & 20.4\% &  55.9\% & 89.2\% &  95.9\% \\
\hline              
\end{tabular}

\newpage

\textbf{Table 4. DGP}= $ 0.5\textsl{Gamma}(4.02804, 0.05576722 ) + 0.5\textsl{log-normal} (4.150614, 0.5214847 )$

\vspace*{0.1cm}

\begin{tabular}{lcccccr}
\hline
   n             &  & 20      & 40  &   60   &   80  &  90   \\
 \hline
$ \hat{\alpha} $ & & 8.836202 &  8.154637    &   8.011385     & 7.880364   & 7.8554   \\
                 & & (3.143420) &  (1.916379)  & (1.491180)  & (1.337800) &  (1.1732)  \\
                 &   &           &          &          &             &          \\
$ \hat{\eta} $  & &  0.122711 & 0.113053   &  0.110831   &  0.108976 &  0.1086  \\
                & & (0.046264) &  (0.028233)   & (0.021856)&  (0.019985) &   ( 0.01761)    \\
                 &   &           &          &          &             &          \\
$ \hat{\mu} $    & & 4.219796 &  4.215739   &  4.216706    & 4.216613 &   4.2166 \\
                 & & (0.080647) &  (0.059911)   & (0.049240) & (0.042338)&  ( 0.03988)   \\
                  &   &           &          &          &             &          \\
$ \hat{\sigma} $  & &  0.358219 &  0.364973   &   0.365404    &  0.367736 &   0.3671\\
                  & & (0.059884) & ( 0.0 42608)  &   (0.034347)& (0.031361)    & ( 0.02839)    \\
                   &   &           &          &          &             &          \\
$ \hat{D}_{\phi_{1}}^{B} $ & & 0.000046 &   0.000042 & 0.000041 & 0.000041 & 0.000040 \\
                           & & (0.000017)&  (0.000012) &  (0.000010) & (0.000009) & (0.000008)   \\
                            &   &           &          &          &                     \\
$ \hat{D}_{\phi_{2}}^{B} $ & & 0.000049  & 0.000047  & 0.000046 & 0.000045  & 0.000045\\
                          & & (0.000017)  &  (0.000012)  & (0.000010) & (0.000009) &  (0.000008)  \\
                            &   &           &          &          &             &          \\
$ U $ & & -0.566535 &  -0.944703   & -1.421004 & -1.675458  & -1.972893     \\
     &  & (0.000006) &  (0.000004)    & (0.000003) &  (0.000002)   & (0.000002)     \\  
Model selection & Gamma         & 2.2\% & 6.7\% & 22.2\%  & 34.7\% & 51.7\%  \\
based on $ U $  & Indecisive    &  96.4\% & 91.3\% & 76.9\% & 64.4\% & 47.7\% \\
                & log-normal    &  1.4\% & 2.0\% & 0.9\%  &  0.9\% & 0.6\%  \\ 
 \hline              
\end{tabular}

\vspace*{0.2cm}

\textbf{Table 5. DGP}= $ 0.75 \textsl{Gamma}(4.02804, 0.05576722 )+ 0.25\textsl{log-normal} (4.150614, 0.5214847 ) $\\

\begin{tabular}{lcccccr}
\hline
   n             &  & 20      & 40  &   60   &   80  &  200   \\
 \hline
$ \hat{\alpha} $ & & 7.680876 &   6.872101    &   6.768990     & 6.709298   & 6.592283\\
                 & & (2.705339) &  (1.545658)  & (1.257030)  & (1.071147) &  (0.647426)         \\
                 &   &           &          &          &             &          \\
$ \hat{\eta} $  & &  0.107086 & 0.095342   &  0.093907   &  0.093040 &  0.091217\\
                & & (0.039593) &  (0.022479)   & (0.018754)&  (0.015924) &   ( 0.009525)          \\
                 &   &           &          &          &             &          \\
$ \hat{\mu} $    & & 4.204221 &  4.202126   &  4.201923    & 4.201670 &   4.202510\\
                 & & (0.090257) &  (0.065006)   & (0.052709) & (0.046475)&  ( 0.029065)             \\
                  &   &           &          &          &             &          \\
$ \hat{\sigma} $  & &  0.386789 &  0.400571   &   0.401475    &  0.402087 &   0.402912\\
                  & & (0.065110) & ( 0.045933)  &   (0.034347)& (0.033488)    & ( 0.20713)    \\
                   &   &           &          &          &             &          \\
$ \hat{D}_{\phi_{1}}^{B} $ & & 0.000040 &   0.000037 & 0.000036 & 0.000035 & 0.000035 \\
                           & & (0.000025)&  (0.000011) &  (0.00009) & (0.000008) & (0.000005)    \\
                            &   &           &          &          &                     \\
$ \hat{D}_{\phi_{2}}^{B} $ & & 0.000043  & 0.000040  & 0.000040 & 0.000039  & 0.000038\\
                          & & (0.000024)  &  (0.000011)  & (0.000010) & (0.000008) &  (0.00000)    \\
                            &   &           &          &          &             &          \\
$ U $ & & -0.654543 &  -1.296389   & -2.172371 & -2.545976  & -3.875059    \\
     &  & (0.000005) &  (0.000002)    & (0.000002) &  (0.000002)   & (0.000001)               \\              
Model selection & Gamma         & 2.2\%  & 18.6\% & 57.8\%  & 72.1\% & 97.2\% \\
based on $ U $  & Indecisive    & 96.5\% & 80.4\% & 42.1\%  &   27.8\% &   0.0\\
                & Log-normal    &  1.3\% & 1.0\% & 0.1\%    & 0.1\% &  2.8\% \\
\hline               
\end{tabular}

\newpage

The first half of each table gives the average values of the multi-step MLE process estimators $ \hat{\alpha}, \hat{\eta},\hat{\mu} $ and $ \hat{\sigma} $, the Bregman divergence test statistics $ \hat{D}_{\phi_{1}}^{B} $ and $ \hat{D}_{\phi_{2}}^{B} $ and the model selection statistic $ U $. The values in parentheses are standard errors. The second half of each table gives the probability of correct selection (PCS) which is in percentage the number of times our proposed model selection procedure based on $ U $, favors the Gamma model, the log-normal model and indecisive. The tests are conducted at 5\% nominal significance level. In the first two sets of experiments ($ \pi= 0.00 $ and $\pi= 1.00$ ) where one model is correctly specified, we use the labels \textit{correct, incorrect} and \textit{indecisive} when a choice is made. The first halves of \textbf{Tables 1-5} confirm our asymptotic results. They all show that the multi-step MLE process estimators $ \hat{\alpha}, \hat{\eta},\hat{\mu} $ and $ \hat{\sigma} $ converge rapidly to their pseudo-true values in the misspecified cases and to their true values in the correctly specified cases as the sample size increases. The statistics $ \hat{D}_{\phi_{1}}^{B}$ and $ \hat{D}_{\phi_{2}}^{B} $  converge approximately to zero at the rate of $n$, as expected when the models are correctly specified and  when the models are  misspecified. With respect to our $ U  $, it diverges to $ -\infty $ at the approximate rate of $ \sqrt{n} $. In \textbf{Tables 3, 4} and \textbf{5}, we observed a large percentage of incorrect decisions. This is because both models are now incorrectly specified. In contrast, turning to the second halves of  \textbf{Tables 1} and \textbf{2}, we first note that the percentage of correct choices using model selection statistic steadily increases and ultimately converge to  100\%. As a consequence, the probability of correct choice (PCS) based on Monte Carlo simulation is found to be significantly higher in chosing the correct model in this selection procedure based on Bregman divergence. The preceding comments for the second halves of \textbf{Tables 1} and \textbf{2} also apply to the second halves of \textbf{Tables 3} and \textbf{4}. The \textbf{Table 5} also confirms our asymptotics results: as sample size increases, the percentage of rejection of both models steadily decreases but still keeping the highest percentage. In all figures we plot the histogram of datasets and overlay the curves for Gamma and log-normal distributions.  They all (figures) show that these two distributions (Gamma and log-normal distributions) are close and closely approximates the data. This is because these distributions are often interchangeable and commonly used to model certain lifetimes in reliability and survival analysis (Wiens, \cite{38}).

\begin{tabular}{lc}
\includegraphics[scale=0.55]{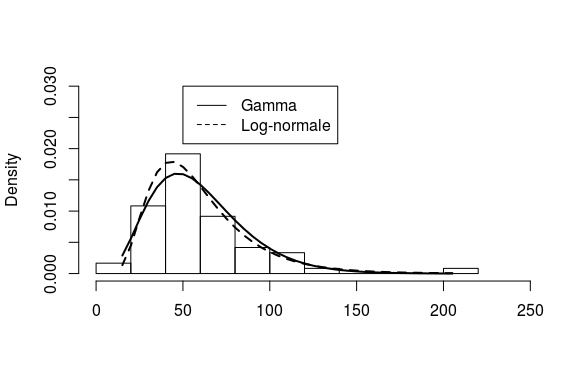}
&
\includegraphics[scale=0.55]{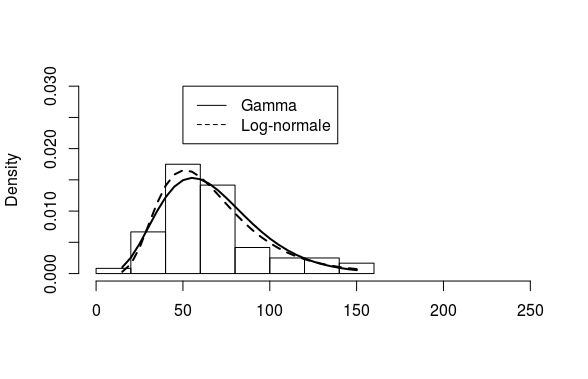}\\
\hspace*{1cm}\textbf{ \underline{Figure 1}. Histogram of DGP } & \hspace*{-2cm} \textbf{\underline{Figure 2}. Histogram of DGP} \\
\hspace*{2.5cm}\textbf{= Log-normal(4.150614, 0.5214847 ),}  & \hspace*{1.5cm} =\textbf{Gamma(4.02804, 0.05576722 ), } \\
\hspace*{2.4cm}\textbf{ with n=60 and $ \pi=0 $.} & \hspace*{0.0cm} \textbf{with  n=60 and $ \pi=1 $. } 
\end{tabular}

\newpage

\begin{tabular}{lc}
\includegraphics[scale=0.55]{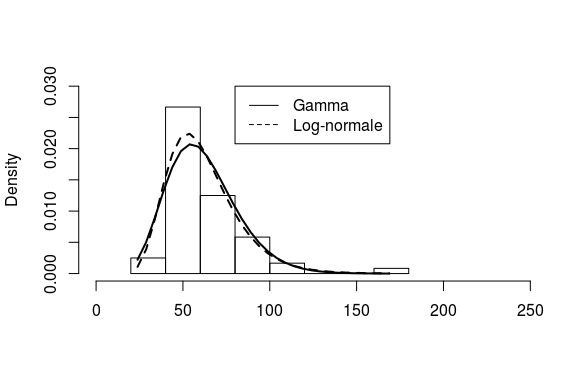}
&
\includegraphics[scale=0.55]{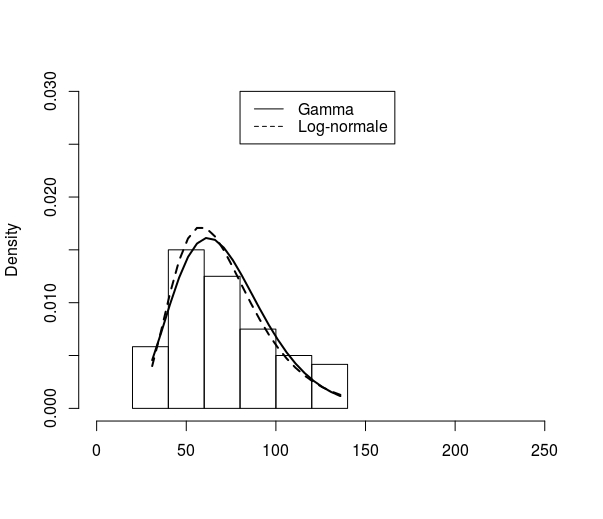}\\
\hspace*{1cm} \textbf{\underline{Figure 3}. Histogram of DGP}     & \hspace*{-2.3cm} \textbf{ \underline{Figure 4}. Histogram of DGP }\\
\hspace*{2.3cm} \textbf{ = 0.25 Gamma(4.02804, 0.05576722 )+}       & \hspace*{2.3cm}\textbf{=  0.5 Gamma(4.02804, 0.05576722 ) + } \\
\hspace*{2.3cm} \textbf{ +0.75 Log-normal(4.150614, 0.5214847 ), }  &\hspace*{2.3cm} \textbf{ +0.5 Log-normal(4.150614, 0.5214847 ),}\\
\hspace*{2.4cm}\textbf{ with n=60, $ \pi=0.25 $.  }                 &    \hspace*{-0.5cm}  \textbf{ with n=60, $ \pi=0.5 $. }
\end{tabular}

\begin{figure}[!h]
\begin{center}
\includegraphics[scale=0.8]{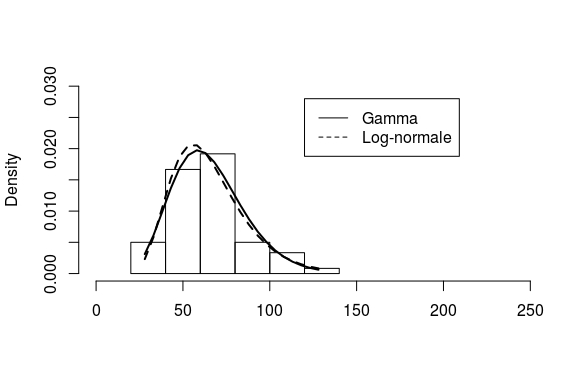}
\end{center}
\end{figure}
\hspace*{3cm} \textbf{\underline{Figure 5}. Histogram of DGP} \textbf{ = 0.75 Gamma(4.02804, 0.05576722 )+}\\
\hspace*{8.2cm} \textbf{ +0.25 Log-normal(4.150614, 0.5214847 ), } \\
\hspace*{8.2cm} \textbf{ with n=60, $ \pi=0.75 $.  } \\  
When the DGP is correctly specified (\textbf{Figure 1}), the log-normal distribution has reasonable chance to be  distinguished from Gamma distribution. Similarly, in \textbf{Figure 2} , as can be seen, the Gamma distribution closely approximates the data sets. In \textbf{Figures
3} and \textbf{5} these two distributions are close but the log-normal
( \textbf{Figure 3} ) and the Gamma distributions ( \textbf{Figure 5} ) does appear to be much closer to the data sets. When $\pi= 0.5$, the distribution for both ( \textbf{Figure 4} ) log-normal distribution and Gamma distribution are nearly similar.

\section{Conclusion}
In this paper we have studied the problem of selecting estimated models using Bregman divergence type statistics. In particular, we have proposed some asymptotically standard normal and hypothesis tests for model selection based on Bregman divergence type statistics that use the corresponding multi-step MLE process estimators. The tests are designed to determine whether the estimated candidate models are as close to the true distribution against alternative hypothesis that one estimated model is closer, where the closeness is measured according to the discrepancy implicit in the Bregman divergence type statistic used. We have established a fundamental property such as the strong consistancy of the Bregman divergence estimator. To facilitate the choice, we have used the bias reduced kernel density estimator to insure the improvement on convergence rate to the true distribution. 
For model selection procedure  based on divergence measures, the Bregman divergence  criterion performs well  especially in small sample.

\end{document}